# Using Computational Grounded Theory to Understand Tutors' Experiences in the Gig Economy

**Lama Alqazlan[1], Rob Procter[1,3] and Michael Castelle[2,3]**
[1]Department of Computer Science, University of Warwick
[2]Centre for Interdisciplinary Methodologies, University of Warwick
[3]The Alan Turing Institute, London, UK
`{Lama.alqazlan|Rob.Procter|M.Castelle.1}`@warwick.ac.uk

## Abstract

The introduction of online marketplace platforms has led to the advent of new forms of flexible, on-demand (or 'gig') work. Yet, most prior research concerning the experience of gig workers examines delivery or crowdsourcing platforms, while the experience of the large numbers of workers who undertake educational labour in the form of tutoring gigs remains understudied. To address this, we use a computational grounded theory approach to analyse tutors' discussions on Reddit. This approach consists of three phases including data exploration, modelling and human-centred interpretation. We use both validation and human evaluation to increase the trustworthiness and reliability of the computational methods. This paper is a work in progress and reports on the first of the three phases of this approach.

## 1 Introduction

The introduction of online platforms has contributed significantly to the changing economic structure and nature of employment (Kenney and Zysman, 2016). In labour markets, platforms are utilised to match and mediate between independent workers and consumers through flexible arrangements, where workers are contracted to perform single discrete tasks and are paid upon task completion, known as the "gig economy" (Broughton et al., 2018; Koutsimpogiorgos et al., 2020). This can provide employment opportunities for those who struggle to find work and the underemployed to supplement their income (Clark, 2021). Conversely, gig employment is also known to be chronically precarious and associated with low remuneration (Duggan et al., 2021; Edward, 2020). To achieve their earnings goals, gig workers have been found to work on average 71 hours, compared to the standard 45 hours per week (Heeks et al., 2021). These hours, however, are not fully paid, as workers spend considerable time waiting on or looking for gigs to perform (Anwar and Graham, 2021). Moreover, in countries such as the UK, gig workers are not eligible for minimum wage, overtime, sick or holiday pay and health insurance primarily because they are classified as "independent contractors" and not employees (Clark, 2021; Heeks et al., 2021), which is currently the focus of legal debates. While some platforms act only as a channel connecting gig workers with customers, others are heavily involved in the process — from job assignments and pricing to work assessment through timing and reviews (Koutsimpogiorgos et al., 2020). This raises doubts as to whether gig workers have sufficient job autonomy to be considered "independent contractors" or if gig economy companies are taking advantage of the current binary worker classification system (i.e., employed or self-employed) to avoid providing employee benefits to employee-like workers (Clark, 2021). Furthermore, the platform economy raises ethical issues relating to their reliance on algorithmic management (Jarrahi et al., 2021; Tan et al., 2021), the lack of transparency on their operation (Jarrahi and Sutherland, 2019), and their inherent power asymmetries, as gig workers are seen as the less powerful party in the process (Koutsimpogiorgos et al., 2020). Moreover, working on some platforms could negatively impact workers' career development, social capital and networks (Duggan et al., 2021).

The issues relating to platform labour are varied and may be more obvious on some platforms than others. Some argue that location-based platforms are less flexible and autonomous than location-independent platforms (Woodcock and Graham, 2019). Working on "microtask" platforms can limit workers' skills and career development (Rani and Furrer, 2019), while workers on "macrotask" platforms feel that their work is underappreciated and underpaid (Nemkova et al., 2019). Many of these issues are especially true for people who rely entirely on the gig economy for their livelihood (Glavin et al., 2021; Koutsimpogiorgos et al., 2020).

Most prior research concerning the experience of gig workers examines platforms such as Uber, Deliveroo and MTurk (Cano et al., 2021; Howcroft and Bergvall-Kåreborn, 2019), while the experiences of the large number of educational labourers who perform tutoring gigs remain understudied. Tutors' experiences are distinct from other gig workers as their tasks typically involve teaching one-to-one sessions which have unique challenges, as these are held in real-time and are of a reasonable length. Therefore, this study aims to contribute to this growing area of research by exploring the experiences of tutors in the gig economy, the problems they face and how their experiences compare to those of other types of gig workers.

## 2 Methodology

### 2.1 Data

The discussion forum and social news aggregator Reddit was manually examined to find relevant subreddits using keywords and platform names, resulting in eighteen subreddits (see Table 1). To retrieve related discussions, we used the Reddit API Wrapper (PRAW), which resulted in approximately 52,000 posts and comments. Then, preprocessing tasks were conducted to convert Reddit's free text into a structure amenable to text mining. This included the removal of stopwords, punctuation, emojis, URLs, lowercasing, lemmatization, and tokenization, resulting in a vocabulary size of 7,491. Samples of posts after the preprocessing steps are shown in Table 2. Finally, before collecting the data, ethics approval was obtained from the Biomedical and Scientific Research Ethics Committee at the University of Warwick.

| Platform-specific Subreddits | Subject-specific and other related Subreddits |
|---|---|
| Vipkid, Preply, MagicEars, Qkids, Cambly, DaDaABC, iTalki, iTutor, Gogokid, Tombac, ZebraEnglish, GoGoKidTeach, Palfish. | OnlineESLTeaching, online_tefl, teachingonline, onlineESLjobs, TeachEnglishOnline. |

Table 1. List of subreddits included in the study

### 2.2 The model

The methodology consists of three main phases.

| Preprocessed tokenized text |
|---|
| ['add','hour','available','asian','company','peak','hour', 'around','west','coast','usa'] |
| ['advice','try','engage','response','move','fair','student', 'learn','time','get','waste'] |
| ['best','case','scenario','mean','lot','new','kid','flock','online', 'long', 'lesson', 'hope', 'put', 'soon', 'right', 'everyone', 'spiral'] |

Table 2. Samples of posts after data preprocessing steps

#### 2.2.1 Phase One: Data Exploration

**(a) Initial data exploration**

Thorough initial data exploration is essential to gain early data-driven insights and to avoid the blind use of unsupervised machine learning algorithms. This is especially important if these are used to replace human reading and judgments on large-scale data, as models output may be varied, misleading or even wrong (DiMaggio, 2015; Grimmer and Stewart, 2013). Thus, we used a constructivist grounded theory (GT) procedure (Charmaz, 2006) to analyse a subset of the data: we randomly selected two small subreddits (GoGoKidTeach and Palfish) with about 160 posts and comments, resulting in a list of themes to be utilised thereafter. GT analysis, however, has its limitations. First, it relies on a series of judgments while coding and interpreting the data, potentially bringing subjectivity into the analysis. Second, the manual reading and analysis of large datasets is onerous and time consuming. These were countered in the next step.

**(b) Validation and further exploration**

To ensure the validity and reliability of the themes identified by GT and expand the data exploration to include the whole corpus, we used Latent Dirichlet Allocation (LDA) (Blei et al., 2003) topic modelling (TM). Labels were produced to describe each topic by examining the 5 highest weighted documents, as close reading of the associated documents can provide sufficient understanding of each topic's essence (Brookes and McEnery, 2019). Topic labels were then compared against GT themes. To assess the validity and reliability of the analysis, one can compare the themes and the topics which emerge from both GT and LDA, a method that Boussalis and Coan (2016) refer to as *concurrent validation*. Furthermore, LDA, with its scalability, accelerates the analysis and may uncover additional topics due to the increased dataset size. The similarity in these two methods allows for validation as both techniques are exploratory, data-driven, iterative and test intermediate versions. While the distinctions between them help to enhance scalability, minimize subjectivity, and reduce the time required for analysis (Baumer et al., 2017).

**(c) Terms extraction**

In this step, we collect the terms for each topic required for Phase Two. First, for topics identified in both GT and LDA or LDA only, the 20 highest weighted terms for each topic in LDA were selected, noting that terms that were judged not to be directly related to the topic were excluded. Second, a list of terms was proposed for topics that appeared only in GT.

### 2.2.2 Phase Two: Modelling — Query-driven topic modelling

In Phase One, the topics of discussions were identified, confirmed, and expanded. Subsequently, query-driven topic modelling (QDTM) (Fang et al., 2021) will be used to model topics from the whole corpus, where the input for the model is a set of query terms for each topic.

The functions and advantages of using this approach are threefold: (1) QDTM can model topics that LDA failed to detect, as LDA ascertains topics using the frequency with which words appear together and thus infrequent topics, regardless of their importance, may not be detected; (2) QDTM uses term expansion techniques, where the input queries are expanded to a set of concept terms using one of three approaches: frequency-based extraction, KL-divergence based extraction, and relevance modelling with word embeddings. This step is particularly helpful, as for most of the topics depend only on terms automatically generated by LDA (with no terms were added to these topics to avoid bias); and (3) QDTM feeds these concept terms into a two-phase framework based on a variation of a Hierarchical Dirichlet Process (HDP) to form the main topic and subtopics. This way of structuring topics is superior to traditional TM (i.e., LDA) which only mines monolayer topics and fails to discover the hierarchical relationship among topics and so is considered an effective way to organise and navigate large-scale data (Dumais and Chen, 2000; Johnson, 1967). This helps to guide analysis and provide the desired level of detail for Phase Three. Finally, the aim is to conduct a human evaluation of topic coherence and exclude non-useful topics.

### 2.2.3 Phase Three: Human-Centred Interpretation — Computational Grounded Theory (CGT)

By this stage, a list of computationally identified, confirmed and human-evaluated topics will have been obtained, each represented by their highest-weighted documents. To perform CGT, samples of the documents will be read in detail and analysed using the conventional GT process. This will add interpretation to the analysis to better understand the topics and assist in the development of higher-level conclusions. The use of TM was essential to discover and classify topics, as the data is too large to be manually read and analysed accurately and efficiently. It also helped to reduce the subjectivity that may come from detailed reading in traditional data analysis methods as a researcher may assign more weight to topics that corroborate their pre-held opinions (Morse, 2015; Nelson, 2017). Finally, as GT involves moving back and forth between the results of the analysis and the data, CGT will take a similar approach as the researcher will return to the data via a structured qualitative analysis after having identified and confirmed the topics (Nelson, 2017).

## 3 Results

### 3.1 GT Analysis

The final high-level themes from Phase One step **(a)** Initial data exploration are listed in Table 3. See Appendix A for a description of each theme.

| Theme no. | Label |
|---|---|
| Theme 1 | Hiring process |
| Theme 2 | New contracts |
| Theme 3 | Job requirements |
| Theme 4 | How tutors consider the job |
| Theme 5 | The class and the students |
| Theme 6 | Teaching material and methods |
| Theme 7 | Bookings and working hours |
| Theme 8 | Payment |
| Theme 9 | Rating system |
| Theme 10 | Reasons to join or leave a platform |
| Theme 11 | COVID-19-related discussions |
| Theme 12 | Technical problems |
| Theme 13 | Miscommunication with platform management |
| Theme 14* | Expressing feelings and sharing experiences |
| Theme 15* | Seeking and providing help and advice |

Table 3. GT themes

*****Note:** Themes 14 and 15, which reflect the underlying purpose of the posts, are excluded from the comparison with LDA as they were generated from observation of the data and due to their abstract nature LDA is not expected to model them.

### 3.2 TM Results

This section presents the results of Phase One step **(b) Validation and further exploration**. This includes the process of finding the optimal LDA model and the final results of the two models in terms of comparing topic labels to GT themes.

At first, the number of topics, K, was assigned the value 13 (see Appendix B, Table B1). Comparing topics labels to GT results, the model found nine topics that correlated with GT themes and one new topic regarding bank transfers and transaction fees. Nonetheless, topics were missing compared to the GT themes, namely: a) reasons to join or leave a platform, b) COVID-19 related discussions, c) teaching material and methods, and d) miscommunication with platform management. Therefore, increasing the number of topics should improve our model. However, since the process of iteratively changing the number of topics and evaluating the results can be time-consuming and impractical, we used the *Tmtoolkit* Python package to compute and evaluate several models in parallel using state-of-the-art theoretical approaches, with the topic range set to between 5 and 30 (see Figure 1).

Here, the optimal number of topics is the one that minimises the average cosine distance between every pair of topics (Cao et al., 2009), and has minimal divergence within a topic (Arun et al., 2010). It is also the one that maximises the word association between pairs of frequent words in each topic (Mimno et al., 2011) and maximises the coherence c_v measure, which calculates the similarity between every top word vector and the sum of all top word vectors (Röder et al., 2015). Therefore, since there was no point or range in the graph where all (or most) measures converged on their maximum or minimum point, the criterion to determine the number of topics was to find a point where all metrics had good values. On this basis, 17 topics were eventually selected.

The 17-topic LDA (see Appendix B, Table B2) was able to model ten topics that correlated with GT themes, where three of them were not in the 13-topic LDA, namely: a) reasons to join or leave a platform, b) miscommunication with platform management and c) teaching material and methods. However, there were still topics missing from the 17-topic LDA: a) how tutors consider the job, b) COVID-19-related discussions and c) discussions around new contracts.

In summary, the two LDA models were collectively able to detect twelve topics identified by GT analysis. However, the models failed to model one topic (COVID-19-related discussions) that was present in GT themes. Conversely, the only topic that was clearly modelled in both LDA models but not in GT was about the issue of bank transfers and transaction fees. Finally, since the aim of this step was to validate and further explore the data, and only one new topic was found even after the number of topics was increased to 17, the aim of this step was fulfilled, and it was not necessary to examine further models. Thus, the final output of this step is 14 topics that will next be considered in QDTM.

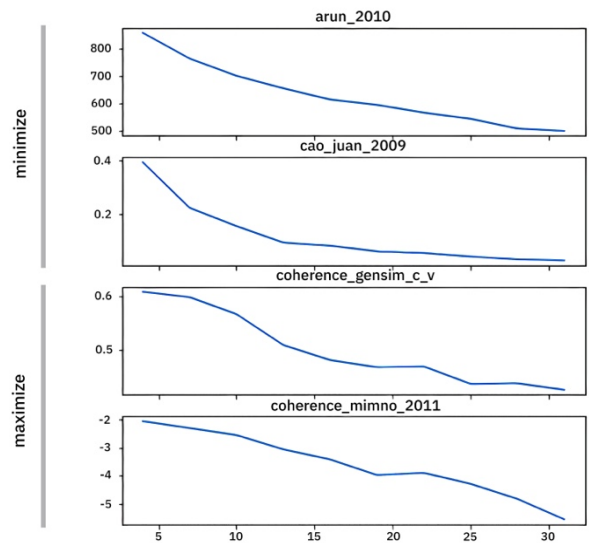

Figure 1. Evaluation of different LDA models when varying the number of topics K.

**Topic Labelling – an explanatory example.**

By way of illustration, this section describes the process of labelling LDA topics as explained in Phase One step (b). Taking topic1 (Appendix B, Table B2) as an example, in order to assign one or more labels to this topic, we began with a close reading of each of the five documents produced by LDA for this topic. It turned out that all the discussions in these documents (posts) revolved around one main topic. For example, in post 1, a tutor talked about their current schedule: "*I open 2 afternoon slots (Mon-Thurs) and 2 morning slots (Wed&Fri&Sat), so that's 13 in total.*". This individual also added information on when tutors can start to get a more steady schedule: "*once you convert the trials, the students will be on your schedule on a weekly basis, so you won't have to worry about bookings*". Similarly, in post 2, another tutor stated: "*I am doing 4-6 priority hours a day, and I have quite a few bookings this week*". Likewise, in post 3, a tutor shared their working hours and the way they schedule bookings: "*I work similar hours (15 hrs.) each week, so do my reservation schedule for a couple of weeks ahead*".

In the remaining two posts, the tutors mentioned days and times of the year with better or worse booking conditions. For instance, post 4 stated: "*Weekends always busy, but lately it's been*

*quiet*", and the tutor who published post 5 tried to explain why: "*It is the last week of school for many kids. I had a lot of cancellations from regulars. My worst day was Monday*".

It is apparent that all the discussions in this topic revolve around tutors' current booking status, their schedule and working hours, and bookings situation at different times. Thus, after understanding each post's content, we decided to label the topic as "Bookings and working hours" which we believe reflects the underlying discussion.

### 3.3 Term extraction results

The terms extracted for each topic that are needed for QDTM are presented in Appendix C, Table C1. As discussed earlier, the topic labels (shown in the first column) are derived from a close reading of topic documents. Then, the 20 most highly weighted terms for each topic (in both LDA models) were examined, and based on our understanding of the topic at hand, we selected the terms that we considered to be most relevant.

The first eight rows in Table C1 represent the topics that have been identified in both LDA models. Following the topic label column, the lists of terms are categorised according to whether they commonly appeared in both 13-topic and 17-topic LDA models, and subsequent columns show terms that appeared uniquely in either 13-topic or 17-topic LDA. Consider the topic "Hiring process" as an example, the terms "interview and apply" appeared in both LDA models, while the terms "referral, link, process and code" only appeared in 13-topic LDA, and the terms "email, profile and application" were uniquely appeared in 17-topic LDA.

The following five rows represent the topics found in only one of the LDA models as well as the terms extracted from them. While, in the last row, we propose terms for the topic "COVID-19-related discussions" since this topic does not appear in either of the LDA models.

## 4 Discussion and conclusions

In this section, we discuss the preliminary results from Phase One (GT and LDA analysis), which have allowed us to gain some early insights toward answering the research question. Therefore, these initial results should be interpreted with caution.

The data analysis showed that tutors in the gig economy seem to experience both platform-related and teaching-related problems. Platform issues include lack of bookings, poor pay, the opacity and the unfairness of rating systems, technical issues, and challenges in reaching the platform management. Comparatively, teaching-related issues appear to start from the first steps in joining these platforms, since platforms offering educational services tend to have a strict hiring process and job requirements. Other issues were found related to the unpaid time spent preparing lessons, as well as the time spent waiting for tutoring gigs to perform. Although tutors can depend on scheduled lessons to help them save time, it may limit their job autonomy and makes it more like a traditional work arrangement. Furthermore, the initial findings suggest that there are other challenges related to teaching that tutors may need to deal with, such as managing class time, dealing with students' different abilities and needs, and fulfilling student expectations. It seems that these teaching-related issues are distinctive of tutors' experiences in the gig economy, which shed light on some aspects of the second research question regarding how their experiences compare to those of other types of gig workers.

In addition to that, to perform tutoring gigs, tutors typically teach for quite long periods of uninterrupted focus, which might pose a barrier to entry for people with caregiving responsibilities or those who lack a quiet place to teach. This does not seem to conform to the assumption that location-independent platforms offer more freedom for workers to perform tasks around life activities. Nonetheless, the initial findings suggest that being location-independent has encouraged many people to join tutoring platforms during COVID-19 lockdowns, most likely due to an increase in free time or because of lost work or unemployment. Tutoring gigs, like most macrotasks, can help workers develop their skills and advance their careers. Furthermore, there is a possibility that tutoring platforms can help people without teaching experience explore the job's suitability. Tutoring online appears to be similar to working for location-dependent platforms in allowing workers to build interpersonal relationships and giving opportunities to develop longer-term relationships with students.

Finally, as only a small portion of the data has been analysed, our understanding of tutors' experience in the gig economy is still limited; we expect to gain more insights and a deeper understanding following our analysis of the data after the completion of phases 2 and 3 of the research plan.

# Appendices

**Appendix** A. Grounded Theory Analysis – A brief description of the themes

1. **Hiring process:** Begins with an application form, then recording a demonstration class, passing a quiz and an interview. Tutors stated that the process is vague and changes frequently. Furthermore, applicants complained of not being informed of their progress, delays in results, and a lack of feedback after rejection.
2. **New contracts:** New tutors and tutors wanting to renew their contracts discussed contracts and related issues, such as increased emphasis on the rating system and pay cuts.
3. **Job requirements:** Being a native speaker is sometimes the only requirement. However, most platforms that provide educational services have stricter requirements, including holding a degree, teaching certificate and experience, and being legally eligible to work. Nonetheless, platforms' acceptance of tutors seems to depend mainly on their need for tutors at the time of applying.
4. **How tutors consider the job:** Tutors tend to view online tutoring as a part-time job to supplement their income, since it is impossible to make a living wage doing it.
5. **The class and the students:** Classes usually take the form of 25-minute one-to-one sessions. Students can be young children or adults. Tutors mostly seemed to have positive opinions of their students.
6. **Teaching material and methods:** When describing lessons, tutors discussed the material they taught and methods they used. Tutors are sometimes required to teach platform-provided materials.
7. **Bookings and working hours:** Bookings are arranged either by the platform or the student, who chooses a suitable tutor based on their profile and rating. Struggling to get bookings is a common problem; some suggest the solution is to be available at all times, which can be impractical and fatiguing. The instability in working hours, both daily and throughout the year, cancellations and student 'no-shows' are among the discussed problems.
8. **Payment:** The average payment is between $14 and $25/hr. Tutors can get bonuses for being on time, teaching during peak hours, or short-notice bookings. Related issues are inadequate payments and pay cuts.
9. **Rating system:** Tutors care deeply about their ratings, as they affect their payment and the number of bookings they receive. Another issue is the opacity of the ratings systems.
10. **Reasons to join or leave a platform:** Some important motives include making money, job autonomy, flexibility and suitability. Reasons to leave a platform include a lack of bookings, inadequate payment, and technical problems. Other reasons relate to the curriculum or feelings of boredom. Some tutors work on multiple platforms to overcome these problems.
11. **COVID-19-related discussions:** Tutors discussed the effects of the pandemic and how it encouraged them join these platforms, due to an increase in free time or because they lost their job. Furthermore, tutors reported that are receiving more bookings due to the closure of schools.
12. **Technical problems:** Not being able to log in, app crashing, or being double-booked are some examples. These can be more frustrating when occurring during lessons, as tutors must pause or cancel the class, which may negatively affect their ratings and payments.
13. **Miscommunication with platforms' management:** Long waiting times for the management team to response, or not receiving one at all, leading tutors to rely on Reddit for answers and solutions.
14. **Expressing feelings and sharing experiences:** Tutors tend to use Reddit forums to express their feelings and share their experiences, both positive and negative.
15. **Seeking and providing help and advice:** An important use of Reddit for tutors at different stages is to ask for or provide advice and help. Tutors seem to be honest, empathetic and generally supportive of each other.

# Appendix B. Top-10 terms and topic labels for LDA models

## 1. 13-topic LDA

| Topic. | Top-10 terms | Label* |
|---|---|---|
| Topic1 | student, teacher, lesson, think, want, really, make, know, learn, people | The class and the students |
| Topic2 | class, time, student, cancel, minute, book, take, get, day, week | Technical problems |
| Topic3 | rating, month, student, get, year, new, week, go, class, contract | The new contracts |
| Topic4 | teach, online, english, experience, school, tefl, course, degree, native, year | Job requirements |
| Topic5 | kid, student, level, use, lesson, question, word, time, slide, class | The class and the students |
| Topic6 | hour, time, week, day, work, schedule, book, class, open, slot | Bookings and working hours |
| Topic7 | company, pay, teacher, work, hire, base, rate, esl, low, people | Payments |
| Topic8 | parent, give, student, kid, say, f***, feedback, teacher, bad, star | Rating system/ The class and the students |
| Topic9 | say, know, post, email, see, people, make, use, send, app | Technical problems |
| Topic10 | pay, lesson, tutor, use, account, student, teacher, money, work, make | Bank transfers and transaction fees |
| Topic11 | work, live, job, china, county, people, think, make, time, get | How tutors consider the job |
| Topic12 | issue, problem, try, work, use, demo, internet, hope, test, good | Technical problems |
| Topic13 | class, minute, min, referral, teach, pay, link, per, good, na | Hiring process |

Table B1. Top-10 terms and topic labels for 13-topic LDA

*Note: Topic labels are produced by reading the 5 highest weighted documents for each topic.

## 2. 17-topic LDA

| Topic | Top-10 terms | Label* |
|---|---|---|
| Topic1 | week, hour, day, time, book, class, slot, schedule, open, month | Bookings and working hours |
| Topic2 | teach, online, work, hour, company, pay, class, experience, tefl, time | Payments/Job requirements |
| Topic3 | student, teacher, work, class, give, rating, really, think, month, company | Rating system |
| Topic4 | student, use, lesson, question, word, ask, say, make, learn, conversation | Teaching material and methods |
| Topic5 | company, teacher, job, work, pay, people, make, money, online, think | Reasons to join or leave a platform |
| Topic6 | class, minute, student, call, time, show, start, late, happen, reservation | Technical problems/ Bookings and working hours |
| Topic7 | class, teacher, parent, time, student, contract, leave, cancel, take, kid | Bookings and working hours |
| Topic8 | lesson, pay, time, student, hour, teacher, rate, tutor, base, minute | Payments |
| Topic9 | kid, teach, level, well, old, year, think, really, feel, say | The class and the students |
| Topic10 | know, anyone, video, tutor, help, ask, let, please, apply, interview | Hiring process / Miscommunication with platform management |
| Topic11 | email, send, message, say, get, group, reply, back, try, see | Hiring process |
| Topic12 | good, student, make, bad, star, sound, say, give, wow, know | The class and the students |
| Topic13 | people, post, think, say, know, f***, mean, name, much, comment | Random |
| Topic14 | english, native, live, country, speaker, language, china, work, american, non | Job requirements |
| Topic15 | month, pay, use, app, work, get, th, phone, tax, year | Payments/ Technical problems |
| Topic16 | feedback, parent, review, keep, student, want, know, class, courseware, see | Rating system |
| Topic17 | rating, account, pay, demo, use, bank, test, paypal, payment, internet | Bank transfers and transaction fees |

Table B2. Top-10 terms and topic labels for 17-topic LDA

*Note: Topic labels are produced by reading the 5 highest weighted documents for each topic.

# Appendix C. Terms per topic required to apply QDTM

| | Topic label | Common terms in both 13- and 17-topic LDA | Terms unique to 13-topic LDA | Terms unique to 17-topic LDA | |
|---|---|---|---|---|---|
| 1 | Hiring process | interview, apply | referral, link, process, code | email, profile, application | Topics appeared in both 13- and 17-topic LDA. |
| 2 | Job requirements | experience, native, degree, tefl, esl, course, company | certificate | country, speaker, language, live, hire, require | |
| 3 | The class and the students | kid, student, level, lesson, class, time, call, teach | video, slide, read, conversation | child, late, show, start, camera, wait, young | |
| 4 | Bookings and working hours | schedule, class, book, slot, hour, time, week, day, month, open, weekend, booking | - | leave, cancel, bonus, trial, regular, ph, cancelation | |
| 5 | Payments | rate, base, pay, low, make | hire, high, offer | price, tax, per | |
| 6 | Rating system | rating, give, feedback, review, bad | star | parent, comment, assessment, good | |
| 7 | Technical problems | app, computer | issue, problem, try, test, connection, internet, email, send, post, check | camera | |
| 8 | Bank transfers and transaction fees | bank, account, pay, paypal, payment | money, platform, price, charge | transfer, payoneer, fee | The new topic, absent from GT |
| 9 | The new contracts | N/A | contract, rating, new, change, year, start | N/A | Topics appeared only in either 13- or 17-topic LDA |
| 10 | How tutors consider this job | N/A | work, live, job, time, money, need, life, income | N/A | |
| 11 | Teaching material and methods | N/A | N/A | use, question, conversation, learn, ask, slide, talk, answer, level, write, read | |
| 12 | Reasons to join or leave a platform | N/A | N/A | job, work, pay, make, money, online, business | |
| 13 | Miscommunication with platform management | N/A | N/A | contact, ticket, response, email, send | |
| | | **Proposed Terms** | | | |
| 14 | COVID-19-related discussions | pandemic, COVID-19, lockdown | | | The missing topic from both LDA models |

Table C1. Terms classified as either common terms generated by both 13- and 17-topic LDA, terms that only appeared in one model or the other, and proposed terms for the missing topic from LDA.